\documentclass[12pt,reqno]{amsart}
\usepackage{amsmath,amsfonts,amsthm,amssymb}
\newcommand\version{October 9, 2007}
\usepackage{amsxtra}

\newtheorem{theorem}{Theorem}[section]
\newtheorem{proposition}[theorem]{Proposition}

\newtheorem{corollary}[theorem]{Corollary}

\theoremstyle{definition}

\newtheorem{example}[theorem]{Example}

\theoremstyle{remark}
\newtheorem{remark}[theorem]{Remark}

\numberwithin{equation}{section}

\newcommand{\C}{\mathbb{C}}
\newcommand{\cl}{{\rm cl}}

\renewcommand{\epsilon}{\varepsilon}

\newcommand{\N}{\mathbb{N}}

\renewcommand{\phi}{\varphi}
\newcommand{\R}{\mathbb{R}}

\DeclareMathOperator{\re}{Re}

\DeclareMathOperator{\Tr}{Tr}
\DeclareMathOperator{\tr}{Tr}

\begin{document}

\title[Matrix-Valued Potentials  --- \version]{Number of Bound
   States of Schr\"odinger Operators with Matrix-Valued Potentials}

\author{Rupert L. Frank}
\address{Rupert L. Frank, Department of Mathematics,
Princeton University, Washington Road, Princeton, NJ 08544, USA}
\email{rlfrank@math.princeton.edu}

\author{Elliott H. Lieb}
\address{Elliott H. Lieb, Departments of Mathematics and Physics,
Princeton University,
      P.~O.~Box 708, Princeton, NJ 08544, USA}
\email{lieb@princeton.edu}

\author{Robert Seiringer}
\address{Robert Seiringer, Department of Physics, Princeton University,
P.~O.~Box 708,
      Princeton, NJ 08544, USA}
\email{rseiring@princeton.edu}

\thanks{\copyright\, 2007 by the authors. This paper may be  
reproduced, in
its entirety, for non-commercial purposes.\\
This work was supported by DAAD grant D/06/49117 (R.F.), by U.S.   
National Science Foundation grants PHY 06 52854
(E.L.) and PHY 06 52356 (R.S.), and by an A.P. Sloan Fellowship (R.S.).}

\begin{abstract} We give a CLR type bound on the number of bound
   states of Schr\"o\-din\-ger operators with matrix-valued potentials
   using the functional integral method of Lieb. This significantly
   improves the constant in this  inequality obtained earlier by
   Hundertmark.
\end{abstract}

\date{\version}

\dedicatory{Dedicated to Jean-Claude Cortet, \\ in appreciation of his  
contribution to Letters in Mathematical Physics}

\maketitle


\section{Introduction}
We consider the Schr\"odinger operator $-\Delta -V(x)$ on $\R^d$, but
with the difference from the usual case that $V$ is a Hermitian
matrix-valued potential. In other words, the Hilbert space is not
$L^2(\R^d)$ but $L^2(\R^d; \C^N)$. The values of functions in this   
space, $\psi
(x) $, are $N-$dimensional vectors. (What we say here easily
generalizes to `operator-valued' potentials, i.e., $\C^N$ is replaced
by a Hilbert space such as $L^2(\R^m)$, but we stay with matrices in
order to avoid technicalities.) The
Cwikel-Lieb-Rozenblum (CLR) bound for $d\geq 3$
in the scalar case $N=1$ states that $\#(-\Delta-V)$, the number of  
negative eigenvalues of
$-\Delta -V$, can be estimated by
\begin{equation}\label{eq:clr}
\#(-\Delta-V) \leq L_{0,d} \int_{\R^d} V_+(x)^{d/2} \,dx\, .
\end{equation}
(Here and below $v_\pm:=(|v|\pm v)/2$ denotes the positive and negative
part of $v$.) We remind the reader that the `semi-classical'   
approximation to $\#(-\Delta-V)$ is given in the scalar case by the   
phase space volume
$$
(2\pi)^{-d} \iint_{\{(p,x)\in\R^d\times\R^d: \, p^2-V(x)<0 \} } dp\,dx
= L_{0,d}^\cl \int_{\R^d} V_+(x)^{d/2} dx
$$
where
\begin{equation*}
     L_{0,d}^{\cl} = (2\pi)^{-d} \int_{\{p\in\R^d:\ p^2<1\}}dp
     = \left( 2^{d} \pi^{d/2} \Gamma(d/2+1) \right)^{-1} .
\end{equation*}
The bound \eqref{eq:clr} was obtained by completely independent  
methods in
\cite{C,L,R}. Later, different proofs were given in \cite{Co,LY}.  
The  best constant, which is close to optimal for $d=3$, was obtained  
in \cite{L} using the  Feynman-Kac formula and Jensen's inequality.

Our goal here is to extend inequality \eqref{eq:clr} to the matrix
case (with a possibly different constant $L_{0,d}$). The motivation
for this extension was the work of Laptev and Weidl \cite{LW1} who
realized that the extension allowed one to conclude that good/sharp  
constants
obtained in low dimensions would automatically give good/sharp  
constants in
higher dimensions.  The fact that the inequality (\ref{eq:clr}) is
valid in the matrix case was proved by Hundertmark \cite{H},
confirming a conjecture in \cite{LW2}. He follows Cwikel's method and
obtains a constant which is far from optimal. Hundertmark points out
that `it would be nice to extend Lieb's [\ldots] proof of the
CLR-bound to operator-valued potentials'. This is the content of this
letter.

\begin{theorem}\label{main}
   Let $d\geq 3$ and assume that $V$ is a function on $\R^d$ taking
values in the Hermitian $N\times N$ matrices. Then
   \begin{equation}\label{eq:main}
     \#(-\Delta -V)
     \leq R_{0,d} \, L_{0,d}^{\cl} \int_{\R^d} \tr_{\C^N} \left[  V_+ 
(x)^{d/2}\right] \,dx
   \end{equation}
   where $R_{0,d}\leq 10.332$ and $V_+:=(|V|+V)/2$.
\end{theorem}

The constant $10.332$ will be obtained for $d=3$ and, by the
Laptev--Weidl method (as used by Hundertmark \cite{H}) it is valid
uniformly for all $d\geq 3$.  We emphasize that our bound on $R_{0,d}$
is slightly worse than the constant $6.87$ in \cite{L} for the
scalar case $N=1$. Still, it improves that of \cite{H} by almost one
order of magnitude. For $d=3$ our bound on $R_{0,3}$ is at most a
factor $2.24$ bigger than the optimal constant in (\ref{eq:main}),
since it is known that $R_{0,3}\geq 8/ \sqrt{3} \approx 4.619$
\cite{LT}.

It is well known that by a simple integration the bound  \eqref 
{eq:main} yields the Lieb-Thirring inequalities
\begin{equation}\label{eq:lt}
   \tr_{L_2(\R^d;\C^N)}\left(-\Delta -V \right)_-^\gamma
   \leq R_{\gamma,d} \, L_{\gamma,d}^{\cl}
   \int_{\R^d} \tr_{\C^N} \left[ V_+(x)^{\gamma+d/2} \right] \,dx
\end{equation}
for all $\gamma>0$, $d\geq 3$ with $R_{\gamma,d}\leq R_{0,d}\leq  
10.332$ and
\begin{equation}\label{eq:class}
   L_{\gamma,d}^{\cl}
   = (2\pi)^{-d} \int_{\R^d} (1-p^2)_+^\gamma \,dp \, .
\end{equation}
Indeed, $R_{\gamma,d}$ is a monotone non-increasing function of $ 
\gamma$ \cite{al}.
Even in the scalar case $N=1$, this yields the best known constants  
in this
inequality for the parameter range $0< \gamma<1/2$. For comparison we  
recall
that the best known bounds for larger values of $\gamma$ are $R_ 
{\gamma,d}\leq
2\pi/\sqrt 3\approx 3.628$ if $\gamma\geq 1/2$ and  $R_{\gamma,d}\leq 
\pi/\sqrt 3\approx
1.814$ if $\gamma\geq 1$ \cite{HLW,DLL}. For $\gamma\geq 3/2$ one has
$R_{\gamma,d}=1$, which is sharp \cite{LW1}. We refer to the surveys   
\cite{H,LW2} for more about inequalities \eqref{eq:lt}.

Apart from yielding very accurate constants we believe that there is a
mathematical interest in extending the path-integral method in \cite 
{L} to the
operator-valued situation. In contrast to the method of \cite{C} used in
\cite{H}. which is rather rigidly based on mapping properties of the  
Fourier
transform, the method of \cite{L} used here works in much wider  
generality,
e.g. on Riemannian manifolds. The only input needed is an upper bound  
on the
heat kernel of the (scalar) unperturbed operator. For example, the
Hardy-Lieb-Thirring bounds in \cite{FLS} extend to the matrix-valued
situation.

As already pointed out, we proceed similarly to \cite{L}. Therefore we
will be brief at some points and ignore some technicalities. There is an
important new ingredient in our proof, however. Since matrices  $W_1, 
\ldots, W_n$ do not commute, in general, we need to work with the   
``time ordering'' of a function $f(\sum_j W_j)$ of their sum. In   
Proposition \ref{jensen} we shall prove a
modification of Jensen's inequality valid in this setting for a certain
class of convex functions $f$.


\section{A trace formula}\label{app:trace}

Given self-adjoint $N\times N$-matrices $W_1,\ldots,W_n$ and a
function $f$ on $\R$, the usual matrix $f(\sum_j W_j)$ is defined by
the spectral projections of $\sum_j W_j$. Instead, we introduce the
``time-ordering'' of the matrix $f(\sum_j W_j)$ as follows. We write
$W_j$ in its spectral representation
$$
W_j= \sum_{k=1}^N w_k^{(j)} P_k^{(j)}\, ,
$$
where $w_k^{(j)}$ are the eigenvalues and $P_k^{(j)}$ the corresponding
orthogonal projections, and define
\begin{equation}\label{eq:deft}
\mathcal T f(W_1,\ldots,W_n) :=
\sum_{k_1,\ldots,k_n=1}^N f\left(\sum_{j=1}^n w_{k_l}^{(j)} \right)
P_{k_1}^{(1)} \cdots P_{k_n}^{(n)}\, .
\end{equation}
Intuitively, this means that when calculating $f(\sum_j W_j)$, one  
puts  all the
$W_1$'s left of the $W_2$'s, the $W_2$'s left of the $W_3$'s, and so  
on, without worrying about
commutators. It is instructive to look at some examples.

\begin{example}\label{ex:polynomial}
   If $f(\mu)=\mu^k$, $k\in\N$, then the definition immediately implies
   \begin{equation*}
     \mathcal T f(W_1,\ldots,W_n)
     = \sum_{j_1+\ldots+j_n=k} \frac{k!}{j_1! \cdots j_n!}
     W_1^{j_1}\cdots W_n^{j_n}\, .
   \end{equation*}
\end{example}

\begin{example}\label{ex:exponential}
   If $f(\mu)=e^{\alpha\mu}$, $\alpha\in\R$, then again by the  
definition (\ref{eq:deft})
   \begin{equation*}
     \mathcal T f(W_1,\ldots,W_n)
     = e^{\alpha W_1}\cdots e^{\alpha W_n}\, .
   \end{equation*}
Similarly, one shows that if $f(\mu)=\mu e^{\alpha\mu}$,
$\alpha\in\R$, then
   \begin{align*}
     & \mathcal T f(W_1,\ldots,W_n)\\
     & \quad = W_1 e^{\alpha W_1} e^{\alpha W_2} \cdots e^{\alpha W_n}
     + e^{\alpha W_1} W_2 e^{\alpha W_2 }\cdots e^{\alpha W_n}
     + \ldots + e^{\alpha W_1} e^{\alpha W_2 }\cdots W_n e^{\alpha  
W_n} \, .
   \end{align*}
\end{example}

We have introduced the notion of time-ordering in order to  
generalize  the trace formula in \cite{L}, which is the starting  
point of the  analysis  leading to \eqref{eq:clr}.

\begin{proposition}\label{trace}
   Let $f$ be a non-negative, lower semi-continuous function $f$ with
   $f(0)=0$, and let
   \begin{equation}\label{eq:f}
     F(\lambda):=\int_0^\infty f(\mu) e^{-\mu/\lambda}\mu^{-1}\,d\mu,
     \qquad \lambda >0.
   \end{equation}
   Then for any sufficiently regular and decaying functions $V$ on $ 
\R^d$, $d\geq 3$, taking values in the non-negative $N\times N$- 
matrices, one has
   \begin{equation}\label{eq:traceformula}
      \begin{split}
        & \tr_{L_2(\R^d;\C^N)}  F(V^{1/2}(-\Delta)^{-1} V^{1/2}) \\
        & \qquad = \int_0^\infty \frac{dt}t  \lim_{n\to\infty}
        \int_{\R^d}\cdots\int_{\R^d} \, dx_1 \cdots dx_n \\
        & \quad\qquad\qquad\qquad\qquad  \prod_{j=1}^n k\left(x_j,x_ 
{j-1},\frac tn\right)
        \tr_{\C^N} \!\left[\mathcal T f\left(\frac tn V(x_1), \ldots,  
\frac tn
V(x_n)\right) \right]
      \end{split}
    \end{equation}
    with the convention that $x_0=x_n$.
\end{proposition}

In the limit $n\to\infty$ the multiple integral on the right side of
\eqref{eq:traceformula} converges to a Wiener integral (the
Feynman-Kac integral); in fact, the right side of
(\ref{eq:traceformula}) is the Trotter product approximation to this
integral \cite{I,RS,simon}.

\begin{proof}
   By an approximation argument \cite[Thm.~8.2]{simon} it suffices  
to  prove this formula for
   \begin{equation*}
     F(\lambda)=\lambda/(1+\alpha\lambda)\, ,
     \quad\quad
     f(\mu)=\mu e^{-\alpha \mu}\,,
   \end{equation*}
   where $\alpha>0$ is a constant. Using the resolvent identity and
   Trotter's product formula, one easily verifies that in this case
   \begin{equation*}
     \begin{split}
       F(V^{1/2}(-\Delta)^{-1}V^{1/2})
       & = V^{1/2}(-\Delta+\alpha V)^{-1}V^{1/2} \\
       & = \int_0^\infty
       V^{1/2}\exp(-t(-\Delta+\alpha V)) V^{1/2}\,dt \\
       & = \int_0^\infty \lim_{n\to\infty} T_n(t) \,dt \, .
     \end{split}
   \end{equation*}
   Here,
   \begin{equation*}
     T_n(t) :=
      V^{1/2}
      \big(\exp(t\Delta/n)\exp(-t\alpha V/n)\big)^n
      V^{1/2}.
    \end{equation*}
    The latter is an integral operator and we evaluate its trace by
integrating its kernel on the diagonal. Let $k$ denote the heat kernel
   $$
   k(x,y,t) := (4\pi t)^{-d/2} \exp(|x-y|^2/(4t))\,.
   $$
Then
   \begin{align*}
     & \tr_{L_2(\R^d;\C^N)} T_n(t)\\
     & \quad = \idotsint \,dx_1 \cdots dx_n
     \prod_{j=1}^n k\left(x_j,x_{j-1},\frac tn\right)
     \tr_{\C^N} \left[ e^{-\frac{\alpha t}n V(x_1)} \cdots
e^{-\frac{\alpha t}n V(x_n)} V(x_n) \right]
     \,.
   \end{align*}
   Cyclical relabeling of the variables leads to
   \begin{align*}
     & \tr_{L_2(\R^d;\C^N)} T_n(t) \\
     & \quad = \frac1t \idotsint \, dx_1 \cdots dx_n\,
     \prod_{j=1}^n k\left(x_j,x_{j-1},\frac tn\right) \tr_{\C^N} \left[
\mathcal T f( t V(x_1)/n, \ldots,t V(x_n)/n) \right]
   \end{align*}
(compare with Example \ref{ex:exponential}).
The claimed formula \eqref{eq:traceformula} follows if one
interchanges the trace with the $t$-integration and the $n$-limit.
\end{proof}


\section{Jensen's inequality and time ordering}

To apply \eqref{eq:traceformula} we need to estimate the trace of a   
time-ordered sum. Recall that
Jensen's inequality says that $\tr f(\sum W_j) \leq n^{-1} \sum \tr f 
(nW_j)$
for $f$ convex. The analog for the time-ordered case, and a certain  
class of $f$'s,  is

\begin{proposition}\label{jensen}
   Assume that
   \begin{equation}\label{eq:jensenass}
     f(\mu)
     = \sum_{j=0}^{\infty} \alpha_j \mu^j + \int_\R e^{-\alpha\mu}
\,d\mu(\alpha)
   \end{equation}
   for some $\alpha_0$, $\alpha_1\in\R$, $\alpha_j\geq 0$ for $j\geq   
2$ and a non-negative measure $\mu$.
Then for any non-negative $N\times N$-matrices $W_1,\ldots,W_n$
   \begin{equation*}
   \re  \tr_{\C^N} \left[\mathcal T f(W_1,\ldots, W_n)\right]
     \leq \frac 1n \sum_{j=1}^n \tr_{\C^N} f(n W_j)\, .
   \end{equation*}
\end{proposition}

Note that the $f$ in (\ref{eq:jensenass}) is convex. We do not know  
whether the statement is
true  for an arbitrary convex function. If it were, the constant in
Theorem \ref{main} could be improved, as explained at the end of this  
letter.

\begin{proof}
   By linearity of the trace it suffices to consider the cases
$f(\mu)=\mu^k$, $k\in\N$, and $f(\mu)= e^{\alpha\mu}$. In the former  
case,
one has by H\"older's inequality for traces (see, e.g., \cite[Thm.  
2.8]{S})
   \begin{align*}
    \re \tr_{\C^N} \left[\mathcal T f(W_1,\ldots, W_n)\right]
     & = \sum_{j_1+\ldots+j_n=k} \frac{k!}{j_1! \cdots j_n!}
     \re \tr_{\C^N} \left[ W_1^{j_1}\cdots W_n^{j_n} \right]\\
     & \leq \sum_{j_1+\ldots+j_n=k} \frac{k!}{j_1! \cdots j_n!}
     \left(\tr_{\C^N} W_1^{k}\right)^{j_1/k} \cdots
     \left(\tr_{\C^N} W_n^{k}\right)^{j_n/k}\\
     & = f\left( \sum_{j=1}^n \left(\tr_{\C^N} W_j^{k}\right)^{1/k}  
\right)\, ,
   \end{align*}
   and the assertion follows from the convexity of $f$. In the latter
case, one has similarly by H\"older's inequality and the
geometric-arithmetic mean inequality
   \begin{align*}
     \re \tr_{\C^N}  \left[\mathcal T f(W_1,\ldots, W_n)\right]
     & = \re \tr_{\C^N} \left[ e^{\alpha W_1}\cdots e^{\alpha
W_n} \right] \\
     & \leq \left(\tr_{\C^N} e^{\alpha n W_1} \right)^{1/n} \cdots
     \left(\tr_{\C^N} e^{\alpha n W_n} \right)^{1/n}\\
     & \leq \frac 1n \sum_{j=1}^n \tr_{\C^N} e^{n \alpha W_j} \, ,
   \end{align*}
   as claimed.
\end{proof}

\begin{corollary}\label{traceestimate}
Assume that $f$ is a non-negative function of the form considered in
Proposition \ref{jensen} and let $F$ be as in \eqref{eq:f}. Then for
any sufficiently regular and decaying function $V$ on $\R^d$ taking
values in the non-negative $N\times N$-matrices, one has
   \begin{equation}\label{eq:traceestimate}
     \tr_{L_2(\R^d;\C^N)} F(V^{1/2}(-\Delta)^{-1} V^{1/2})
     \leq \frac{1}{(4\pi)^{d/2}}
     \left(\int_0^\infty \frac {f(s)}{s^{d/2}} \frac{ds}s \right)
     \int_{\R^d} \tr_{\C^N} \left[V(x)^{d/2}\right] \, dx\, .
    \end{equation}
\end{corollary}

\begin{proof}
   Combining Proposition \ref{jensen} with Proposition \ref{trace} we  
obtain
   \begin{equation*}
     \begin{split}
       & \tr_{L_2(\R^d;\C^N)} F(V^{1/2}(-\Delta)^{-1} V^{1/2}) \\
       & \leq \int_0^\infty \frac{dt}t  \lim_{n\to\infty}
       \int_{\R^d}\cdots\int_{\R^d}
       \prod_{j=1}^n k\left(x_j,x_{j-1},\frac tn\right)
       \frac 1n \sum_{j=1}^n \tr_{\C^N} f(t V(x_j))
       \, dx_1 \cdots dx_n \, .
      \end{split}
    \end{equation*}
(Here we have used that the left side of (\ref{eq:traceformula}) is   
real, hence only
the real part of $\Tr \mathcal T f$ contributes to the integral.)
    The semi-group property implies
    \begin{align*}
      & \frac 1n \int_{\R^d}\cdots\int_{\R^d}
      \prod_{j=1}^n k\left(x_j,x_{j-1},\frac tn\right)
      \sum_{j=1}^n \tr_{\C^N} f(t V(x_j)) \, dx_1 \cdots dx_n \\
      & \qquad = \frac 1n \sum_{j=1}^n \int_{\R^d} k\left(x_j,x_{j},t 
\right)
      \tr_{\C^N} f(t V(x_j)) \, dx_j
      = \frac 1{(4\pi t)^{d/2}} \int_{\R^d} \tr_{\C^N} f(t V(x)) \, dx 
\, .
    \end{align*}
    Denoting the eigenvalues of $V(x)$ by $v_1(x)\leq \ldots\leq v_N(x)$
one finds that
    \begin{align*}
      \int_0^\infty \frac{dt}t \frac{\tr_{\C^N} f(t V(x))}{t^{d/2}}
      = \sum_{j=1}^N \int_0^\infty \frac{dt}t \frac{f(t v_j(x))}{t^{d/ 
2}}
      = \sum_{j=1}^N v_j(x)^{d/2} \int_0^\infty \frac{ds}s \frac
{f(s)}{s^{d/2}}\, ,
    \end{align*}
thereby   proving the assertion.
\end{proof}


\section{Proof of Theorem \ref{main}}

First we assume that $d=3$. By the variational principle we can assume
that $V(x)$ is a non-negative matrix for all $x$, and by an  
approximation
argument we can assume that $V$ is smooth and rapidly decaying. For
any increasing function $F$ on $(0,\infty)$ the Birman-Schwinger
principle implies that
\begin{equation}\label{eq:bs}
   \#(-\Delta-V) \leq F(1)^{-1} \tr_{L^2(\R^3;\C^N)} F(V^{1/2}(- 
\Delta)^{-1} V^{1/2})\, .
\end{equation}
We choose $F=F_a$ of the form \eqref{eq:f} where $a>0$ is a parameter  
and
$f=f_a$ is defined by
$$
f_a(\mu)=\frac{\mu^2}{\mu+a} = \mu -a +\frac{a^2}{\mu+a}
= \mu - a+ a^2 \int_0^\infty e^{-t(\mu+a)}\,dt\, .
$$
Since this function is of the form considered in Proposition  \ref 
{jensen} we can
apply Corollary~\ref{traceestimate} and get in view of \eqref{eq:bs}
\begin{equation*}
   \#(-\Delta-V) \leq C_a \int_{\R^3} \tr_{\C^N} \left[V(x)^{3/2} 
\right] \, dx
\end{equation*}
where
\begin{align*}
   C_a & := (4\pi)^{-3/2} F_a(1)^{-1}
   \left(\int_0^\infty \frac {f_a(s)}{s^{3/2}} \frac{ds}s \right) \\
   & =  \frac 18 (\pi a)^{-1/2}
   \left( 1 + a e^{a} \int_a^\infty  e^{-s}\frac{ds}s
\right)^{-1} .
\end{align*}
The result follows by choosing $a=1.13$, which approximately
minimizes $C_a$.

Now we assume that $d\geq 4$. We will use the Laptev-Weidl strategy  
to  reduce this case to the case
$d=3$ as in \cite{H}. We note that by a straightforward
approximation argument as in \cite{LW1} the inequality for $d=3$ holds
also for $N=\infty$, i.e., if $V(x)$ assumes values in the compact
self-adjoint operators on a separable Hilbert space. Introduce
variables $x=(x_1,x_2)\in \R^d$ where $x_1\in\R^3$ and
$x_2\in\R^{d-3}$. We decompose the Laplacian correspondingly as
$-\Delta=-\Delta_1-\Delta_2$ and define, for fixed $x_1\in\R^3$,
$W(x_1):=(-\Delta_1-V(x_1,\cdot))_-$. If $V$ is, say, smooth
with compact support, then $W(x_1)$ is a compact
operator in $L^2(\R^{d-3},\C^N)$  for every $x_1$. The variational  
principle and the
inequality for $d=3$ imply that
\begin{equation*}
\#(-\Delta-V) \leq \#(-\Delta_1 - W)
\leq R \, L_{0,3}^{\cl} \int_{\R^3} \tr_{L^2(\R^{d-3},\C^N)}
\left[W(x_1)^{3/2}\right]\,dx_1 \,.
\end{equation*}
By the result of Laptev and Weidl \cite{LW1}, one has
\begin{equation*}
\tr_{L^2(\R^{d-3},\C^N)} \left[ W(x_1)^{3/2} \right]
\leq L_{3/2,d-3}^{\cl} \int_{\R^{d-3}} \tr_{\C^N} \left[V(x_1,x_2)^{d/ 
2}\right]\,dx_2
\end{equation*}
with the constant $L_{3/2,d-3}^{\cl}$ from \eqref{eq:class}. Noting that
$L_{0,3}^{\cl}L_{3/2,d-3}^{\cl} = L_{0,d}^{\cl}$ we obtain the  
assertion of
Theorem \ref{main}.

\begin{remark}
If the estimate in Proposition \ref{jensen} held for all convex   
functions (not merely for those of the form \eqref{eq:jensenass}),   
then we could choose $f_a(\mu)=(\mu-a)_+$ in the preceding proof, as   
in \cite{L}, and would get the same constant as in the scalar case.
\end{remark}


\bibliographystyle{amsalpha}

\begin{thebibliography}{HLW}
\bibitem[AL]{al}  M. Aizenman, E.H. Lieb, {\it On Semi-Classical
Bounds for Eigenvalues
of Schr\"odinger Operators}, Phys. Lett. {\bf 66A} (1978), 427--429.

\bibitem[Co]{Co}
   J.~G.~Conlon, \textit{A new proof of the Cwikel-Lieb-Rosenbljum   
bound}. Rocky Mountain J. Math. \textbf{15} (1985), no. 1, 117--122.
\bibitem[C]{C}
M.~Cwikel, \textit{Weak type estimates for singular values and the  
number of
bound states of Schr\"odinger operators}. Ann. Math. \textbf{106}  
(1977),
93--102.
\bibitem[DLL]{DLL}
J.~Dolbeault, A.~Laptev, M.~Loss, \textit{Lieb-Thirring inequalities  
with
improved constants}. J. Eur. Math. Soc., to appear. Preprint:
arXiv:0708.1165v2 [math.AP].
\bibitem[FLS]{FLS}
   R. L. Frank, E. H. Lieb and R. Seiringer, \textit{Hardy-Lieb-Thirring
inequalities for fractional Schr\"odinger operators}. J. Amer. Math.  
Soc., to
appear. Preprint: arXiv:math/0610593v2 [math.SP]
\bibitem[H]{H}
   D.~Hundertmark, \textit{On the number of bound states for Schr 
\"odinger
operators with operator-valued potentials}. Ark. Mat. \textbf{40}  
(2002),
73--87.
\bibitem[HLW]{HLW}
D. Hundertmark, A. Laptev and T. Weidl, \textit{New bounds on the
Lieb-Thirring constants}. Invent. Math., \textbf{40} (2000), 693--704.
\bibitem[I]{I}
T. Ichinose, \textit{Norm convergence of the Trotter product formula  
for Schr\"odinger operators via the Feynman-Kac formula}. Path  
integrals: Dubna '96,  341--346, Joint Inst. Nuclear Res., Dubna, 1996.
\bibitem[LW1]{LW1}
A.~Laptev, T.~Weidl, \textit{Sharp Lieb-Thirring inequalities in high
dimensions}. Acta Math. \textbf{184} (2000), 87--111.
\bibitem[LW2]{LW2}
A. Laptev, T. Weidl, \textit{Recent results on Lieb-Thirring  
inequalities}.
Journ\'ees ``\'Equations aux D\'eriv\'ees Partielles" (La Chapelle sur
Erdre, 2000), Exp. No. XX, Univ. Nantes, Nantes, 2000.
\bibitem[LY]{LY}
   P. Li, S. T. Yau, \textit{On the Schršdinger equation and the   
eigenvalue problem}. Comm. Math. Phys. \textbf{88} (1983), no. 3,   
309--318.
\bibitem[L]{L}
E. H. Lieb, {\it Bounds on the eigenvalues of the Laplace and
    Schr\"odinger operators}, Bull. Amer. Math. Soc. {\bf 82} (1976),
751--752. \textit{The number of bound states of one body Schr\"odinger
operators and the Weyl problem}. Proc. A.M.S. Symp. Pure Math. \textbf 
{36}
(1980), 241--252.
\bibitem[LT]{LT} E.~H.~Lieb, W.~Thirring, \textit{Inequalities for the
     moments of the eigenvalues of the Schr\"odinger Hamiltonian and
     their relation to Sobolev inequalities}. Studies in Mathematical
   Physics, 269--303. Princeton University Press, Princeton, NJ, 1976.
\bibitem[RS]{RS}
    G. V. Rozenblum, M. Solomyak, \textit{The Cwikel-Lieb-Rozenblyum  
estimator for generators of positive semigroups and semigroups  
dominated by positive semigroups}. St. Petersburg Math.~J.~\textbf{9}  
(1998),  no. 6, 1195--1211.
\bibitem[R]{R}
   G.~V.~Rozenblum,
   \textit{Distribution of the discrete spectrum of singular
     differential operators}.
   Soviet Math. Dokl. \textbf{13} (1972), 245--249, and
   Soviet Math. (Iz. VUZ) \textbf{20} (1976), 63--71.
\bibitem[S1]{S} B. Simon, {\it Trace ideals and their
       applications}, Second edition, Mathematical Surveys and  
Monographs
     {\bf 120}, American Mathematical Society, Providence, RI, 2005.
\bibitem[S2]{simon} B. Simon, {\it Functional Integration and Quantum
      Physics}, Second edition, Amer. Math. Soc., Providence, RI, 2005.



\end{thebibliography}

\end{document}